\documentclass[12pt, a4paper, fleqn]{article}
\usepackage{latexsym,amsmath,amsfonts,amssymb}
\usepackage{graphicx}
\usepackage{indentfirst}
\usepackage{cite}
\usepackage[cp1251]{inputenc}

\begin{document}

\begin{center}
 {\Large \bf Kinetics of the elementary act \\ of electrochemical reactions \\ at the semiconductor--electrolyte solution\\ interface}

\medskip

{\bf Sergii Kovalenko $^{\dag \ddag}$} {\bf and  Veniamin Soloviev $^\ddag$}
\\
{\it $^\dag$~Institute of Mathematics, Ukrainian National Academy of Sciences,
\\
3 Tereshchenkivs'ka Str., Kyiv 01601, Ukraine}
\\
{\it $^{\ddag}$~Division of Physics, Poltava National Technical Yuri Kondratyuk University,
\\
24 Pershotravnevyi Prosp., Poltava 36011, Ukraine}

\medskip

E-mail: kovalenko@imath.kiev.ua

\medskip

\end{center}

\begin{abstract}
In the framework of the quantum-mechanical theory of elementary act of non-adiabatic electrochemical reactions, it is carried out the calculation of the discharge current of ions at the semiconductor--electrolyte solution interface using the model of isotropic spherically symmetric band. It is shown that our results generalize the well-known formulae for the current density obtained by Dogonadze, Kuznetsov, and Chizmadzhev [R.\,R. Dogonadze, A.\,M. Kuznetsov, and Yu.\,A. Chizmadzhev, The kinetics of some heterogeneous reactions at semiconductor--electrolyte interface, Zhur. Fiz. Khim. 38 (1964) 1195--1202]. The average densities of states in the valence band and the conduction band of the semiconductor electrode in the heterogeneous charge transfer are found.
\end{abstract}

\textbf{Key words}: elementary act of electrochemical reactions, quantum-mechanical theory, density of states.

\section{Introduction}

One of the modern theories of elementary act of charge transfer at the solid--polar liquid interface is the quantum-mechanical theory, whose main statements were proposed by Dogonadze, Chizmadzhev and Kuznetsov in the first half of the 60's of the 20th century \cite{dog62-1,dog62-2,dog63,dog64,dog65,kuz64}(see, also, \cite{krish79}). In the last few decades, the efforts of researchers working in this theory aimed both at improving the well-known theoretical principles and the development of new theoretical concepts including heterogeneous proton and other heavy ions transfer at different interfaces, theoretical modelling heterogeneous processes with new electrode materials as high-temperature superconductors and nanotubes, etc. (see, for instance, \cite{schmic99,kuz99,kuz00}).

At the same time, it should be noted that the quantum-mechanical theory of heterogeneous charge transfer at the semiconductor--electrolyte solution interface, which was created by Dogonadze et al. in the 60's of the last century \cite{dog63,dog64,dog65,kuz64} has not been investigated for the more. Note that the main outcomes of this theory coincide, in general, with the statements of the semi-phenomenological theory of elementary act of electrochemical reactions developed by Gerischer in the early sixties of the last century (see, for instance, \cite{schmic10,sato98}). At the same time, it should be stressed that the existing theories only qualitatively describe the electrochemical processes on semiconductor (insulator) electrodes.

Note that within the existing quantum-mechanical theory of elementary act of non-adiabatic charge transfer at the semiconductor--electrolyte solution interface, the calculation of the discharge currents of ions was carried out under some conditions and simplifications. The most significant of which are
\begin{itemize}
    \item[1)] neglecting the real geometry of ions discharged at the electrode (the model of points charges);
    \item[2)] input assumptions about the absence of specific adsorption of ions on the electrode;
    \item[3)] assuming that the discharge of ions occurs at a distance as close as possible to the electrode, i.e. from the Helmholtz layer surface;
    \item[4)] the gas of free charge carriers in the semiconductor or insulator electrode is not degenerate (the Maxwell--Boltzman statistics);
    \item[5)] assuming that the density $\rho = \rho(E)$ of states in the semiconductor weakly depends on the energy, i.e., actually, the model $\rho(E) = \rho = \mbox{const}$ is used, while the constant $\rho$ is a parameter of the theory.
\end{itemize}

In this article, we will calculate the discharge currents of ions on the semiconductor electrode abandoning the last assumption. However, for the density of states in the conduction band and the valence band we will use the standard law for the density of states near the bottom of the isotropic spherically symmetric band. Also, for the discharge currents will be obtained accurate analytical expressions as opposed to the earlier works, where only some asymptotic expressions were obtained for certain additional restrictions on the parameters available in the theory.

\section{Kinetics of the elementary act of electron transfer at the semiconductor electrode}

Let us consider the elementary act of discharge of an ion involved in the electrolyte solution at the surface of a semiconductor electrode:
\begin{equation}\label{d1}
OX^{z+} + e \rightarrow RED^{(z-1)+}.
\end{equation}
In accordance with the general theory (see, for example, \cite{dog64}), the current density of the reaction can be written in the form
\begin{equation}\label{d2}
j = j_{\mathrm{ns}} - j_{\mathrm{sn}} = \left(j_{\mathrm{ns}}^{(e)} + j_{\mathrm{ns}}^{(p)} \right) - \left(j_{\mathrm{sn}}^{(e)} + j_{\mathrm{sn}}^{(p)} \right),
\end{equation}
where $j_{\mathrm{ns}}$ and $j_{\mathrm{sn}}$ are the anode and the cathode densities of current, respectively; the upper indexes $e$ and $p$ define the type of band, namely, $e$ and $p$ are used for the conduction band and the valence band, respectively.

The relation between the cathode $j_{\mathrm{sn}}$ and the anode $j_{\mathrm{ns}}$ currents is as follows:
\begin{equation}\label{d3}
j_{\mathrm{ns}} = j_{\mathrm{sn}} \cdot e^{\frac{e\eta}{kT}},
\end{equation}
where $\eta$ is the overvoltage in the bulk of electrode, $k$ is Boltzmann's constant.

Note that the last formula is valid for both the hole $j^{(p)}$ and the electronic $j^{(e)}$ components of the density of discharge current, i.e.
\begin{equation}\label{d4}
j^{(p)}_{\mathrm{ns}} = j^{(p)}_{\mathrm{sn}} \cdot e^{\frac{e\eta}{kT}},
\end{equation}
\begin{equation}\label{d5}
j^{(e)}_{\mathrm{ns}} = j^{(e)}_{\mathrm{sn}} \cdot e^{\frac{e\eta}{kT}}.
\end{equation}

Hereafter, in order to simplify the cumbersome mathematical formulae, we assume that the deviations from equilibrium in the electrode is small and neglecting the potential drop in the electrolyte diffusion layer. Then, taking into account the conditions formulated in the introduction, the expressions for the cathode currents $j_{\mathrm{sn}}^{(p)}$ and $j_{\mathrm{sn}}^{(e)}$ can be written in the following form \cite{dog64, kuz00}:
\begin{equation}\label{1}
j_{\mathrm{sn}}^{(p)} = e c_{\mathrm{ox}} l_{\mathrm{ef}} \int_{-\infty}^{E_p} \rho_p(E) W(E,\eta) dE,
\end{equation}
\begin{equation}\label{2}
j_{\mathrm{sn}}^{(e)} = e c_{\mathrm{ox}} l_{\mathrm{ef}} \int_{E_e}^{+\infty} \rho_e(E) W(E,\eta) \exp\left(-\frac{E-E_F}{kT} \right) dE,
\end{equation}
where $c_{\mathrm{ox}}$ and $c_{\mathrm{red}}$ are the concentrations of the oxidized and the reduced forms of the ion in the bulk of the electrolyte solution, respectively, $l_{\mathrm{ef}}$ is the effective thickness of the reaction region, $E_F$ is the Fermi level, $E_p = E_F - \Delta_p + e(\varphi_n - \varphi_k)$, $E_e = E_F + \Delta_e + e(\varphi_n - \varphi_k)$, $\varphi_n$ and $\varphi_k$ are the potentials in the bulk of electrode and at the contact with the electrolyte solution, respectively, $\Delta_e$ and $\Delta_p$ are the gaps between $E_F$ and the lower edge of the conduction band and the upper edge of the valence band, respectively. $W(E,\eta)$ is the rate constant for the electron transfer from the level $E$, at the overvoltage $\eta$, which can be read as follows: \cite{dog62-1}
\begin{equation}\label{d6}
W(E,\eta) = \left(\frac{\pi}{\hbar^2\lambda kT} \right)^{\frac{1}{2}}|L_{\mathrm{sf}}|^2 \exp\left\{-\frac{(\lambda + \Delta G^0(E,\eta))^2}{4\lambda kT} \right\},
\end{equation}
where $\lambda$ is the total environmental and local classical reorganization Gibbs free energy, in the limit of linear electronic-vibrational coupling, $L_{\mathrm{sf}}$ is the electron exchange factor, which is assuming to be constant, $\Delta G^0(E, \eta) = e\eta + kT \ln\frac{c_{\mathrm{ox}}}{c_{\mathrm{red}}} - (E - E_F)$ is the driving force relating to the electronic energy $E$, $\rho_p(E)$ and $\rho_e(E)$ are the electronic densities of states in the valence band and the conduction band of the electrode, respectively.

As it follows from formulae \eqref{1} and \eqref{2}, the dependence of the discharge currents from the bands structure of the electrode is mainly determined by the characteristics of the lower edge of the conduction band and the upper edge of the valence band. From the band theory is known that for semiconductors, particularly with a narrow band gap, this structure can be quite complicated \cite{aske85,shali85}. However, the first works on the quantum-mechanical theory of elementary act of non-adiabatic electrochemical reactions at the semiconductor electrode \cite{dog63,dog64,kuz64} assumed that the function $\rho(E)$ weakly depends on energy, so it can be taken outside the integral sign, i.e. the model $\rho(E) = \rho = \mbox{const}$ was used. Note that the constant $\rho$ was used as a parameter of the theory without specifying its value. Later, suggesting that the concentrations of electrons and holes, respectively, in the conduction band and the valence band are small, and using the standard model of the isotropic spherically symmetric band
\begin{equation}\label{3}
\rho_p(E) = \frac{m^{\ast}_p}{2\pi^2\hbar^3}\sqrt{2m^{\ast}_p(E_p - E)},
\end{equation}
\begin{equation}\label{4}
\rho_e(E) = \frac{m^{\ast}_e}{2\pi^2\hbar^3}\sqrt{2m^{\ast}_e(E - E_e)},
\end{equation}
where $m^{\ast}_p$ and $m^{\ast}_e$ are the effective masses of electrons and holes near the edge of the relevant band, Dogonadze and Kuznetsov \cite{dog75} obtained an approximate estimation for the energy level $E^{\ast}$ of the most probable electron transfer. Namely, they established that the approximate equalities $E^{\ast} - E_e \sim kT$ for the conduction band and $E_p - E^{\ast} \sim kT$ for the valence band, respectively, take place. Note, that Gerischer has also obtained the similar result within the framework of the semi-phenomenological theory \cite{sato98}.

Taking into account relations \eqref{3} and \eqref{4} for the densities of states, we have calculated the discharge currents $j_{\mathrm{sn}}^{(p)}$ and $j_{\mathrm{sn}}^{(e)}$. In this case, the integrals in formulae \eqref{1} and \eqref{2} can be calculated accurately without putting any additional assumptions unlike the approach of Dogonadze and his collaborators. Without going into complicated technical calculations, we present only the final result
\begin{equation}\label{5}
j_{\mathrm{sn}}^{(p)} = j_0^{(p)} \exp\left\{-\beta_p \cdot \frac{e\eta_k}{kT} \right\},
\end{equation}
\begin{equation}\label{6}
j_{\mathrm{sn}}^{(e)} = j_0^{(e)} \exp\left\{(1-\beta_e) \cdot \frac{e\eta_k}{kT} - \frac{e\eta}{kT} \right\},
\end{equation}
where $j_0^{(p)}$ and $j_0^{(e)}$ are the hole and electron exchange currents, which are respectively
\begin{equation}\label{d7}
j_0^{(p)} = A_p N_p \exp\left\{-\frac{(\lambda + \Delta_p - e(\varphi^0_n - \varphi^0_k))^2}{8\lambda kT}\right\} D_{-\frac{3}{2}} \left[\frac{\lambda + \Delta_p - e(\varphi^0_n - \varphi^0_k)}{\sqrt{2\lambda kT}} \right],
\end{equation}
\begin{equation}\label{d8}
j_0^{(e)} = A_e N_e \exp\left\{-\frac{(\lambda + \Delta_e + e(\varphi^0_n - \varphi^0_k))^2}{8\lambda kT}\right\} D_{-\frac{3}{2}} \left[\frac{\lambda + \Delta_e + e(\varphi^0_n - \varphi^0_k)}{\sqrt{2\lambda kT}} \right],
\end{equation}
where $A_{p,e} = \frac{el_{\mathrm{ef}}}{\hbar} |L_{\mathrm{sf}}|^2 \left(\frac{\pi}{\lambda kT} \right)^{\frac{1}{2}} \left(\frac{2\lambda}{kT} \right)^{\frac{3}{4}} c_{\mathrm{red}}^{\beta_{p,e}} c_{\mathrm{ox}}^{1-\beta_{p,e}}$, $N_p$ and $N_e$ are the effective densities of states in the valence band and the conduction band, respectively, $\varphi^0_n$ and $\varphi^0_k$ are the equilibrium potentials in the bulk of the semiconductor electrode and at the contact with the electrolyte solution, respectively, $\eta_k = \varphi_k - \varphi_k^0$ is the overvoltage at the contact with the electrolyte solution, $D_{-\frac{3}{2}}(x)$ is the Weber--Hermite function \cite{bat53}.

In \eqref{5} and \eqref{6} the constants $\beta_p$ and $\beta_e$, which are the coefficients of proportionality between the change in the activation energy and the heat of reaction, i.e., the factors that go into the Br\"{o}nsted relation, are as follows:
\begin{equation}\label{d9}
\beta_p = \frac{1}{2} + \frac{\Delta_p - e(\varphi^0_n - \varphi^0_k)}{2\lambda}, \quad \beta_e = \frac{1}{2} - \frac{\Delta_e + e(\varphi^0_n - \varphi^0_k)}{2\lambda}.
\end{equation}
Formulae \eqref{5} and \eqref{6} are the general expressions for the discharge currents of ions at the semiconductor--electrolyte solution interface. At the same time, under some physically motivated assumptions, the expressions for the exchange currents $j_0^{(p)}$ and $j_0^{(e)}$ can be further simplified. Since the analysis is similar for both the electron and the hole exchange currents, we restrict ourselves to the current $j_0^{(p)}$.

First, we note that for the majority of electrochemical reactions, the total reorganization energy of the system has a great value ($\sim 10 \, \mbox{eV}$), therefore the relation $\lambda \gg kT$ takes place. This means that the condition
\begin{equation}\label{d10}
\frac{\lambda + \Delta_p - e(\varphi^0_n - \varphi^0_k)}{\sqrt{2\lambda kT}} \gg 1
\end{equation}
holds. Using the well-known expansion of the Weber--Hermite function $D_{-\frac{3}{2}}(z)$ in the asymptotic series \cite{bat53}
\begin{equation}\label{d11}
D_{-\frac{3}{2}}(z) = z^{-\frac{3}{2}} \cdot e^{-\frac{z^2}{4}} + O(z^{-2}), \ z \rightarrow + \infty,
\end{equation}
where $z = \frac{\lambda + \Delta_p - e(\varphi^0_n - \varphi^0_k)}{\sqrt{2\lambda kT}}$, after the corresponding calculations one obtains
\begin{equation}\label{7}
j_0^{(p)} \approx \frac{el_{\mathrm{ef}}}{\hbar} |L_{\mathrm{sf}}|^2 \left(\frac{8\pi}{kT} \right)^{\frac{1}{2}} \frac{\lambda N_p c_{\mathrm{red}}^{\beta_{p}} c_{\mathrm{ox}}^{1-\beta_{p}}}{(\lambda + \Delta_p - e(\varphi^0_n - \varphi^0_k))^{\frac{3}{2}}} \, \exp\left\{-\frac{(\lambda + \Delta_p - e(\varphi^0_n - \varphi^0_k))^2}{4\lambda kT}\right\}.
\end{equation}

Expression \eqref{7} cannot be future simplified, for example, in the case of a semiconductor electrode with the wide forbidden band, when $\lambda \sim \Delta_p$. In the case $\lambda \gg \Delta_p$, for the value $\beta_p$ the approximate equality
\begin{equation}\label{8}
\beta_p = \frac{1}{2} + \frac{\Delta_p - e(\varphi^0_n - \varphi^0_k)}{2\lambda} \approx \frac{1}{2}
\end{equation}
holds. Taking into account \eqref{8}, expression \eqref{7} reads as follows
\begin{equation}\label{9}
j_0^{(p)} \approx \frac{el_{\mathrm{ef}}}{\hbar} |L_{\mathrm{sf}}|^2 \left(\frac{8\pi}{\lambda kT} \right)^{\frac{1}{2}} (c_{\mathrm{ox}} c_{\mathrm{red}})^{\frac{1}{2}} N_p \exp\left\{-\frac{(\lambda + \Delta_p - e(\varphi^0_n - \varphi^0_k))^2}{4\lambda kT}\right\}.
\end{equation}

The assumptions used to obtain formula \eqref{9} are similar to those, which were used previously in \cite{dog63,dog64}. Comparing \eqref{9} with the results obtained in \cite{dog64} (see formula (29)), we obtain the value of the constant $\rho_p$, which by its physical meaning is the average value of states in the valence band of the electrode materials in heterogeneous charge transfer
\begin{equation}\label{10}
\rho_p = \frac{\sqrt 2}{kT} \, N_p.
\end{equation}
Note that in \cite{dog63,dog64} the value of $\rho_p$ has not been obtained. The last formula shows that $\rho_p$ is proportional to the effective density of states in the valence band $N_p$ with the coefficient of proportionality $\frac{\sqrt 2}{kT}$.

Find the value of the energy level $\widetilde{E}$ in the valence band, which corresponds to the obtained value of $\rho_p$, i.e. $\rho_p = \rho_p(\widetilde{E})$. Taking into account \eqref{3}, it is easy to find
\begin{equation}\label{d12}
E_p - \widetilde{E} = \frac{\pi}{2} \, kT \approx 1.6 \, kT.
\end{equation}
We see that the value obtained for $\widetilde{E}$ coincides with the outcomes of Gerischer and Dogonadze et al. that the main contribution to the heterogeneous charge transfer is made by the energy levels, which are separated from the edges of the valence band and the conduction band on the value of order $kT$.

Note that a similar analysis can be carried out in the general case without putting the additional asymptotic assumptions, but the expressions for $\rho_p$ and $\rho_e$ will have a much more complex structure.

It should be stressed that our results are of purely theoretical nature, but it is interested to compare them with experimental data, specifically, with the results of experiments of Shapoval et al. concerning the possibility of drawing on natural and synthetic diamonds the galvanic coating without preliminary depositing a conducting film \cite{kush90, shap95, mal00, mal14}. The authors of the experiments have not proposed a convincing interpretation of the results obtained. There have been made only the phenomenological conclusion that the surface conductivity of diamond in oxide melts arises from the specific electrochemical properties of the diamond--ionic melt interface, specifically, the occurrence of interfacial redox reactions. However, up to date, the question of the nature and mechanism of the surface conductivity is debatable.

From our point of view, the obtained expressions (\ref{5}) and (\ref{6}) for the discharge current of ions at the semiconductor (covalent insulator)--electrolyte solution interface can be applied to explain the possibility of the surface conductivity of insulators in the ionic melts using experimental data of the volts-amperometric and potentiometric studies \cite{shap95, mal00, mal14} in combination with the high-accurate quantum-chemical calculations of such quantities like the energy of reorganization, the transmission coefficient, etc. \cite{naz00, sol04}.

We are going to return to a detailed discussion of these questions in our forthcoming publications.

\section{Conclusions}

In this article, within the quantum-mechanical theory of elementary act of non-adiabatic electrochemical reactions, we carried out the calculation of discharge current of ions on the semiconductor electrode. Our calculations were based on the model of isotropic spherically symmetric band with the root law dependence from the energy of the density of states, in contrast to the earlier works on quantum-mechanical modelling of physical and chemical properties of a solid electrode, where it was not taken into account the energy dependence of the densities of states in the valence band and the conduction band of the electrode. Note that this model adequately describes the features of the band structure of the electrode with the low concentrations of holes and electrons near the edges of the valence band and the conduction band, respectively.

	The main result of the paper is formulae (\ref{5}) and (\ref{6}) for the hole $j_{\mathrm{sn}}^{(p)}$ and  the electron  $j_{\mathrm{sn}}^{(e)}$ components of the cathode current flowing in the studied system. Comparison of these formulae with those obtained previously in [3, 4] shows that formally they coincide. However, the expressions for the corresponding exchange currents $j_0^{(p)}$ and $j_0^{(e)}$, that are the parts of formulae (\ref{5}) and (\ref{6}), are significantly different from those previously obtained by Dogonadze et al. for the semiconductor-electrolyte solution interface.

	In the asymptotic approximation $\lambda \gg kT$ and $\lambda \gg \Delta_{p,e}$, we compared our results with those of Dogonadze et al. [4]. It was shown that the average densities of states in the valence band and the conduction band of the electrode in heterogeneous charge transfer within the used model of isotropic spherically symmetric band are proportional to the effective densities of states in the relevant bands with the coefficient of proportionality $\frac{\sqrt 2}{kT}$.

	Subsequently, the theory developed in this work will be extended to the case of an electrode with the degenerate gas of free charge carriers using more sophisticated models of the structure of the valence and the conduction bands.


\begin{thebibliography}{99}

\bibitem{dog62-1} R.\,R. Dogonadze and Yu.\,A. Chizmadzhev, Dokl. Akad. Nauk SSSR, Ser. Fiz. Khim. \textbf{144}, 1077 (1962).

\bibitem{dog62-2} R.\,R. Dogonadze and Yu.\,A. Chizmadzhev, Dokl. Akad. Nauk SSSR, Ser. Fiz. Khim. \textbf{145}, 849 (1962).

\bibitem{dog63} R.\,R. Dogonadze and Yu.\,A. Chizmadzhev, Dokl. Akad. Nauk SSSR, Ser. Fiz. Khim. \textbf{150}, 333 (1963).

\bibitem{dog64} R.\,R. Dogonadze, A.\,M. Kuznetsov and Yu.\,A. Chizmadzhev, Zhur. Fiz. Khim. \textbf{38}, 1195 (1964).

\bibitem{kuz64} A.\,M. Kuznetsov and R.\,R. Dogonadze, Izv. Akad. Nauk SSSR, Ser. Khim. \textbf{12}, 2140 (1964).

\bibitem{dog65} R.\,R. Dogonadze and A.\,M. Kuznetsov, Elektrokhimiya \textbf{1}, 742 (1965).

\bibitem{krish79} L.\,I. Krishtalik, Electrode Reactions and the Mechanism of Elementary Act, Nauka, Moskow 1979 (in Russian).

\bibitem{kuz99} A.\,M. Kuznetsov and J. Ulstrup, Electron Transfer in Chemistry and Biology. An Introduction to the Theory, John Wiley \& Sons, Chichester 1999.

\bibitem{schmic99} W. Schmickler, Annu. Rep. Prog. Chem., Sect. C: Phys. Chem. \textbf{95}, 117 (1999).

\bibitem{kuz00} A.\,M. Kuznetsov and J. Ulstrup, Electrochim. Acta \textbf{45}, 2339 (2000).

\bibitem{sato98} N. Sato, Electrochemistry at Metal and Semiconductor Electrodes, Elsevier, Amsterdam 1998.

\bibitem{schmic10} W. Schmickler and E. Santos, Interfacial Electrochemistry, Springer, Berlin 2010.

\bibitem{aske85} B.\,M. Askerov, Phenomena of Electrons Transfer in Semiconductors, Nauka, Moskow 1985 (in Russian).

\bibitem{shali85} K.\,V. Shalimova, Physics of Semiconductors, Energoatomizdat, Moskow 1985 (in Russian).

\bibitem{dog75} R.\,R. Dogonadze and A.\,M. Kuznetsov, Prog. Surf. Sci. \textbf{6}, 1 (1975).

\bibitem{bat53} H. Bateman and A. Erd\'{e}lyi, Higher Transcendential Functions, Vol. 2, McGraw-Hill Book Company Inc., New York 1953.

\bibitem{kush90} H.\,B. Kushkhov, V.\,I. Shapoval, and A.\,N. Baraboshkin, Dokl. Akad. Nauk SSSR, Ser. Fiz. Khim. \textbf{312}, 1405 (1990).

\bibitem{shap95} V.\,I. Shapoval, I.\,A. Novosyolova, V.\,V. Malyshev, and H.\,B. Kushkhov, Electrochim. Acta \textbf{40}, 1031 (1995).

\bibitem{mal00} V.\,V. Malyshev, I.\,A. Novosyolova, A.\,I. Gab, A\,D. Pisanenko, and V.\,I. Shapoval, Theor. Found. Chem. Eng. \textbf{34}, 391 (2000).

\bibitem{mal14} V.\,V. Malyshev, A.\,I. Gab, A\,D. Pisanenko, V.\,V. Soloviev, and L.\,A. Chernenko, Mat.-wiss. Werkstofftech. \textbf{45}, 51 (2014).

\bibitem{naz00} R.\,R Nazmutdinov, G.\,A. Tsirlina, O.\,A. Petrii, Yu.\,I. Kharkats and A.\,M. Kuznetsov, Electrochim. Acta \textbf{45}, 3521 (2000).

\bibitem{sol04} V.\,V. Solov'ev, V.\,V. Malyshev, and A.\,I. Gab, Theor. Found. Chem. Eng. \textbf{38}, 205 (2004).

\end{thebibliography}
\end{document}